\documentclass[10pt,aps,pre,twocolumn,notitlepage,superscriptaddress,preprintnumbers]{revtex4-1}

\usepackage{amsmath,amssymb,amsfonts}
\usepackage{graphicx}           
\usepackage{epstopdf}         
\usepackage{xcolor}             
\usepackage[colorlinks=true,linkcolor=blue,citecolor=blue]{hyperref}	
\usepackage{dcolumn}            
\usepackage[FIGTOPCAP]{subfigure}          
\usepackage{float}
\usepackage{hyperref}
\usepackage{url}                
\usepackage[normalem]{ulem}     
\usepackage{accsupp}            
\usepackage{textcomp}
\let\iint\undefined
\let\iiint\undefined
\usepackage{wasysym}

\definecolor{darkcyan}{HTML}{008080}
\definecolor{darkcyan}{rgb}{0.0, 0.55, 0.55}
\definecolor{darkblue}{HTML}{000080}
\definecolor{darkred}{HTML}{800000}
\newcommand\arrows[0]{\ensuremath{\stackrel{\textcolor{darkblue}{\leftarrow}}{\textcolor{darkcyan}{\boldsymbol{\rightarrow}}}}}

\begin{document}

\title{Individual versus Social Benefit on the Heterogeneous Networks}
\author{Z. Babaee}
\affiliation {Department of Physics, Institute for Advanced Studies in 
Basic Sciences (IASBS), Zanjan 45137-66731, Iran}
\author{M. Bagherikalhor}
\email{mahsa.bagherikalhor@gmail.com}
\affiliation{Department of Physics, Shahid Beheshti University, 
Evin, Tehran 19839, Iran}
\author{L. Elyasizad}
\affiliation {Department of Physics, Institute for Advanced Studies in 
Basic Sciences (IASBS), Zanjan 45137-66731, Iran}
\author{M. D. Niry}
\affiliation {Department of Physics, Institute for Advanced Studies in 
Basic Sciences (IASBS), Zanjan 45137-66731, Iran}
\author{G. R. Jafari}
\email{g\_jafari@sbu.ac.ir}
\affiliation{Department of Physics, Shahid Beheshti University, 
	Evin, Tehran 19839, Iran}
\affiliation{
	Irkutsk National Research Technical University,  $664074$, Lermontov Street, $83$, Irkutsk, Russia
	 }

\date{\today}

\begin{abstract}
The focus of structural balance theory is dedicated to social benefits, 
while in a real network individual benefits sometimes get the importance 
as well. In Strauss's model, the local minima are modeled by considering 
an individual term besides a social one and the assumption is based on 
equal strength of individual benefits. The results show that the 
competition between two terms leads to a phase transition between 
individual 
and social benefits and there is a critical point, 
$CP$, that represents a first-order phase transition in the 
network. Concerning a real network of 
relations, individuals adjust the strength of their relationships based on 
the benefits they acquire from. Therefore, addressing heterogeneity in the 
individual interactions, we study a modified version of 
Strauss's model in which the first term represents the heterogeneous  
individual benefit by $\theta_{ij}$, and the coefficient of the second 
term, $\alpha$, measures the strength of social benefit. Our studies show 
that 
there is a region where the triangles are in a crumpled state rather than 
being dispersed in the network and increasing 
the heterogeneity of individual benefits results in the narrower region of 
crumpled state. Out of this region, the network is a mixture of links and 
triangles and the value of $\alpha$ determines whether the individual 
benefit or social benefit overcomes. For the small value of $\alpha$ the 
individual benefit dominates whereas in the large value of $\alpha$ the 
social 
benefit overcomes.\\
\end{abstract}
\maketitle

\section{Introduction}
In 1946 Heider proposed a theory in social psychology, 
so-called structural balance theory. Its assumption is based on friendly 
and enmity relationships, $\{+1,-1\}$. The theory goes beyond pairwise 
interactions and considers triadic groups of 
interaction and addresses social benefits 
\cite{FritzHeider,heider2013psychology,KU_AKOWSKI_2005}. Then, it was 
extended to graph theory by Cartwright and Harary\cite{Cartwright}. 
Investigation of the 
social networks and understanding their structural features have been 
considered in 
\cite{Albert_2002,Dorogovtsev_2002,Newman_2003,newman2002random, 
	jin2001structure,pham2021balance,minh2020effect,Montakhab1,Montakhab2}.
Furthermore, there is a comprehensive literature 
of the structural balance theory in different branches of science from 
mathematical sociology \cite{leik1975mathematical} to ecology \cite{Saiz} 
and biology \cite{moradimanesh2021altered,10.3389/fphys.2020.573732}. A 
bunch of researches in physics focus on social benefits and have looked it 
through the lens of structural balance and represented an analytical 
mean-field solution for the considered networks 
\cite{Freshteh,Mahsa,Razieh,Amir}. 
In general, a significant property of networks that can be particularly 
highlighted in social networks is clustering \cite{watts1998collective} 
which means networks have a heightened density of short loops of various 
lengths \cite{caldarelli2004structure}. Newman showed that clustering can 
have a considerable effect on the large-scale structure of networks 
\cite{newman2003properties}. Holme \textit{et al.} stated that social 
networks typically show a high clustering \cite{holme2002growing}. 
Real social networks indicate that each individual has a local benefit in 
addition to the social one. In 1986, 
Strauss developed a general class of models for interactions.  Strauss's 
model mimics individual(local) benefit besides social(global) one in a 
network. He 
introduced a model that a link of $\{1,0\}$ represents the presence or 
absence of a relationship between two individuals and the interaction 
between the links arise as clustering \cite{strauss1986general}. 

Later on, Burda \textit{et al.} gave a comprehensive discussion of 
Strauss's model and introduced an action that counts the number of 
triangles. 
The extension of the action in terms of all orders showed that the 
perturbation series breaks down when the local benefit is comparable to 
the global one and the condensed phase of triads emerges 
\cite{burda2004network,burda2004perturbing}. Due to the lack of a 
complete solution of the model by using perturbation series, Newman 
presented a mean-field solution to solve the problem analytically. He 
studied the competition between individual and social 
benefit in a random network that links can either exist or not and the 
interaction of links results in a phase in which many triangles forms. He 
showed that there is a complete agreement between the analytical and 
simulation results \cite{Newman2005}. Results from numerical simulation 
show that as a consequence of competition between local and global benefit 
the system experiences a phase transition where triads tend to form a clan 
rather than being spread in the network uniformly.

Several publications have provided a more complete view of networks by 
considering heterogeneity in the strength of the interactions between 
individuals 
\cite{YookWeighted,Nohedgeweights,barrat2004architecture,BarratWeightedEvolving}.
In a real network of relations, each person makes 
different relationships and joins various groups for distinct goals, and
making and ending relationships depends on local or social benefits.
Hence, there is always competition between 
local choices and social benefits. Strauss's 
model \cite{strauss1986general} studies the 
competition between social and individual benefits, based on the 
the assumption that individual relationships are equivalent while in 
a network of relationships, it is clearly not the case, and individuals 
assign different importance to their relations. Now the question is how
the dynamic of the network will change under consideration of 
heterogeneous individual interactions? 

In this paper, we consider heterogeneous local benefits in contrast to 
Strauss's model which all local benefits are the same. Our main purpose is 
to study how heterogeneity of relationships affects the dynamics of the 
network. We use exponential random graphs 
\cite{frank1986markov,billard2005journal,snijders2006new,snijders2011statistical},
which is a powerful tool to study the networks. It is an extension of the 
statistical mechanics of Boltzmann and Gibbs to the networks. We present 
an analytical mean-field solution following the approach of 
\cite{Newman2005,park2004solution,park2004statistical}. 
At last, the analytic arguments were completed by numerical simulations 
and we see the agreement between the theory and the simulation. 

\section{\label{sec:method} Analytical solution under Mean-field method }
Strauss's model is based on the competition between local and 
global benefits \cite{strauss1986general}. The contribution of these two 
elements are illustrated in the first and second terms of the following 
Hamiltonian:
\begin{equation}\label{eq1}
	H(S)=\theta\sum_{i<j}s_{ij}-\alpha\sum_{i<j<k}s_{ij}s_{jk}s_{ki},
\end{equation}
where $s_{ij}$ is the element of the adjacency matrix of the network $G$ 
which contains $N$ nodes. The possible values of $s_{ij}$ are $\{0,1\}$
that indicate the absence and presence of a link between nodes $i$ and 
$j$, respectively. Since the relationships 
are bilateral the adjacency matrix is symmetric, $s_{ij}=s_{ji}$. The 
network $G$ represents a specific configuration from a set of 
possible configurations which could be created by adjusting the elements 
of the adjacency matrix. The first term of the Hamiltonian~(\ref{eq1}) 
shows the individual benefits, an increase in $\theta$ heightens the 
energy of the network which is not desirable so the network gets sparse 
and the density of links decreases. When this term is 
dominant, the creation of a link between two nodes will be expensive; it 
will cause an increase in the total energy of the network resulting in a 
tendency to isolate the nodes. The second term encourages the creation of 
triads for $\alpha>0$. The coefficient $\alpha$ shows the strength of the 
triad relationships. The negative sign indicate that large values of 
$\alpha$ encourage the creation of triads followed by the creation of 
clusters. These two terms compete for links that participate in the 
formation of triads. The network will be sparse for large values of 
$\theta$ and small values of $\alpha$, but as $\alpha$ gets larger triads 
are more likely to form and the number of links in the network will 
increase. In~\cite{Newman2005} the authors by 
representing a mean-field solution for this Hamiltonian found that the 
system experiences a phase transition between the low and high density of 
triads.

Here, we are interested in studying the effect of heterogeneity in the 
Strauss model. We apply non-uniform stochastic values on each link 
between every two nodes (individuals). To do so, we made a modification to 
the Strauss model by adopting a Gaussian distribution for individual 
benefits with a specific mean 
value ($\bar{\theta}$) and variance ($\sigma^2$) for $\theta_{ij}$ 
that assigns random value $\theta_{ij}$ to each link. The values are 
quenched meaning that they are constant on the time scale over which the 
links fluctuate. The introduced Hamiltonian is:
\begin{equation}\label{eq2}
	H(S) = \sum_{i<j} \theta_{ij}s_{ij} - \alpha \sum_{i<j<k} 
	s_{ij}s_{jk}s_{ki}.
\end{equation}

The random value assigned to each link, $\theta_{ij}$, indicates all 
relationships are not equally important. We set $\bar{\theta}$ to a 
specific positive value since every person has some degree of selfishness 
and isolation, and concerning the variance, $\theta_{ij}$'s spread around 
$\bar{\theta}$. The links with large values of $\theta_{ij}$ are more 
probable to be removed than the links with small $\theta_{ij}$. Also, 
negative $\theta_{ij}$'s rarely appear. The negative values of 
$\theta_{ij}$ are desirable since they decrease the energy of the network. 
So these links usually behave like a permanent backbone and participate in 
the creation of the triads. It should be noted that $\bar\theta>0$, either 
in the modified model or in the Strauss model. The ratio of the number of 
positive $\theta_{ij}$s to the number of negative $\theta_{ij}$s depends 
on the values of $\bar{\theta}$ and $\sigma$. The competition between the 
two terms of the Hamiltonian and their coefficients determines the 
probable configurations of the network.

We investigate the effects of different ratio of $\bar{\theta}$/$\sigma^2$ 
on the dynamics of the network and we study network stability as a 
consequence of the competition between local and global benefits. 
The constant coefficient $\alpha$ shows the strength of a triadic 
relationships in the second term of the Hamiltonian. The positive sign of 
first term tells us the network tends to be sparse and in the second term 
as $\alpha$, gets larger triads are more likely to form. The competition 
of these two terms will result in notable behavior of the system that we 
aim to study.
The analytical solution of Eq.~(\ref{eq1}) indicates the occurrence of a 
first order phase transition in the phase space of parameters of the 
Hamiltonian. Here, based on the mean-field approach we present an 
analytical solution for the proposed Hamiltonian Eq.~(\ref{eq2}). Then we 
test the accuracy of the analytic solutions by applying the 
Metropolis-Hastings algorithm in the simulation section. We rewrite the 
Hamiltonian in the following way:
\begin{equation}\label{eq3}
\begin{aligned}
& H = H_{ij} + H', \\
& H_{ij} = \theta_{ij}s_{ij} - \alpha 
s_{ij}{\sum_{k\neq{i,j}}}s_{jk}s_{ki},\\
\end{aligned}
\end{equation}
here $H_{ij}$ consists of all terms in the Hamiltonian, Eq.~(\ref{eq2}), 
that involve $s_{ij}$, $H'$ is related to the remaining terms. Notice, we 
have two independent probability distributions $P(G)$, $P(\theta)$. The 
first one is coming from the exponential random graph method which states 
the probability of selecting a specific graph among the set of available 
configurations in each of which $s_{ij}$ can be either $0$ or $1$. The 
second one is a normal distribution that indicates the probability of 
random values that are being assigned to links. Now, let calculate the 
average of a link, so \\
\begin{equation}\label{eq4}
\begin{aligned}[b]
{\langle s_{ij}\rangle}_{G,\theta} =&{}\sum_{s_{ij}=\{0,1\}} s_{ij} 
P(s_{ij})\\
& = 0\times P(s_{ij}=0) + 1\times P(s_{ij}=1)\\
& = 
{\left\langle\frac{e^{-\theta_{ij}+\alpha\sum_{k\neq{i,j}}s_{jk}s_{ki}}}
{1+e^{-\theta_{ij}+\alpha\sum_{k\neq{i,j}}s_{jk}s_{ki}}}\right\rangle}_{\!\!{G,\theta}}\\
& = 
{\left\langle\frac{1}{e^{\theta_{ij}-\alpha(N-2)q}+1}\right\rangle}_{\!\!\theta}\\
& =  
\frac{1}{\sqrt{2\pi\sigma^2}}\int_{-\infty}^{\infty}\frac{e^{\frac{-(\theta_{ij}-\bar{\theta})^2}{2\sigma^2}}}
{e^{\theta_{ij}-\alpha(N-2)q}+1}
d\theta_{ij},
\end{aligned}
\end{equation}
in the last line we used Gaussian probability distribution for averaging 
over 
$\theta$. We replace $\sum_{k\neq{i,j}}s_{jk}s_{ki}$ in the denominator 
with $(N-2)\left\langle s_{jk}s_{ki}\right\rangle$, and define 
$q\equiv\langle s_{jk}s_{ki}\rangle$ as the average of a two-star (two 
links connected to a common node) in the network.  As Newman has 
explained, the mean-field approach is proper for large networks since, in 
the limit of $N\rightarrow\infty$, this approach is exact 
\cite{Newman2005}. This quantity, $q$, can be interpreted as the local 
field each link feels. We found that average of a link, $p\equiv\langle 
s_{ij}\rangle_{G,\theta}$, is a function 
of some parameters
\begin{equation}\label{eq5}
 p = f(q,\alpha,N,\bar{\theta},\sigma).
\end{equation}

We follow the same approach to derive the average of two-star:
\begin{eqnarray}\label{eq6}
H            & = & H_{jk,ki} + H', \nonumber\\
H_{jk,ki} & = & \theta_{jk}s_{jk} - \alpha s_{jk}{\sum_{l\neq{i,j,k}}}s_{jl}s_{kl}+\theta_{ki}s_{ki} \nonumber\\
               &  &- \alpha s_{ki}{\sum_{l\neq{i,j,k}}}s_{kl}s_{il} - 
               \alpha s_{ij}s_{jk}s_{ki},
\end{eqnarray}
the first part, $H_{jk,ki}$ includes all terms consist of $s_{jk}$, 
$s_{ki}$ and the second term, $H'$, involves the rest. Concerning the 
assumption that all $\theta_{ij}$, $\theta_{jk}$, 
$\theta_{ki}$ are coming from Gaussian probability distributions with the 
same mean and variance, we calculate $q\equiv\langle 
s_{jk}s_{ki}\rangle_{G,\theta}$,

\begin{widetext}
\begin{equation}\label{eq7}
\begin{aligned}
& {\left\langle s_{ij}s_{jk}\right\rangle_{G,\theta}} = 
\left\langle\frac{e^{\alpha 
s_{ij}}}{e^{\theta_{ki}-\alpha\sum_{l\neq{i,j,k}}s_{kl}s_{li}}+e^{\theta_{jk}-
			\alpha\sum_{l\neq{i,j,k}}s_{jl}s_{lk}}+e^{\theta_{jk}-\alpha
			\sum_{l\neq{i,j,k}}s_{jl}s_{lk}+\theta_{ki}-
			\alpha\sum_{l\neq{i,j,k}}s_{kl}s_{li}}+e^{\alpha 
			s_{ij}}}\right\rangle_{\!\!{G,\theta}}\\
& = {\left(\tfrac{1}{\sqrt{2\pi\sigma^2}}\right)}^2 
	\iint_{-\infty}^\infty
\frac{e^{\alpha
		p} \ e^{\frac{-(\theta_{jk}-\bar{\theta})^2}{2\sigma^2}} \ 
		e^{\frac{-(\theta_{ki}-\bar{\theta})^2}{2\sigma^2}}}
{(e^{\theta_{ki}-\alpha(N-3)q}+1)(e^{\theta_{jk}-
		\alpha(N-3)q}+1)+(e^{\alpha p}-1)} 
d\theta_{jk} \ d\theta_{ki}.
\end{aligned}
\end{equation}
Using Eq.~(\ref{eq5}), we have a self-consistence equations for 
the average of two-star, $q$, as:
 \begin{equation}\label{eq8}
 q = g(p,q,\alpha,N,\bar{\theta},\sigma).
 \end{equation}
\end{widetext}

Obtaining the self-consistence equation (\ref{eq8}) is the result of 
considering heterogeneous local benefits and this is the remarkable 
difference of our model with the Strauss's ones.
If variance tends to zero Gaussian probability distribution reduces to 
\textit{Dirac delta} function so, all $\theta_{ij}$'s are approximately 
equal and we obtain Strauss's Hamiltonian. Investigating 
Eqs.~(\ref{eq5},~\ref{eq8}) opens our horizon through the dynamics of the 
network when individual relationships have different importance. Approving 
by simulation results in Sec.~\ref{sec:res}, plotting the average of 
two-star, order parameter, versus coefficient $\alpha$ displays 
the existence of a first-order phase transition due to the sudden jump 
between low and high density of triads in the network.

The last quantity we manage to calculate is the average of triangle, 
$r\equiv\langle s_{ij}s_{jk}s_{ki}\rangle_{G,\theta}$. 
Similar to preceding steps, we divide the Hamiltonian into two different 
parts,
\begin{eqnarray} 
	H &=& H_{ij,jk,ki} + H', \nonumber\\
	H_{ij,jk,ki} &=& \theta_{ij}s_{ij} + \theta_{jk}s_{jk} + 
	\theta_{ki}s_{ki} - \alpha s_{ij}s_{jk}s_{ki}\nonumber\\
	& & -\alpha s_{ij} \sum_{l\neq i,j,k} s_{il}s_{jl} -\alpha s_{jk} 
	\sum_{l\neq i,j,k} s_{jl}s_{kl} \nonumber\\ 
	& & -\alpha s_{ki} \sum_{l\neq i,j,k} s_{kl}s_{il},
	\label{eq9}
\end{eqnarray}
where $H_{ij,jk,ki}$ dedicates the terms related to each link $s_{ij}$, 
$s_{jk}$, and $s_{ki}$ individually and a triad of all those. We calculate 
the average value of triangle:
\begin{widetext}
\begin{equation}
\begin{aligned}
& \langle s_{ij}s_{jk}s_{ki}\rangle_{G,\theta} = 
{\left\langle\frac{e^{\alpha}}{(e^{\theta_{jk}-\alpha(N-3)q}+1)
(e^{\theta_{ki}-\alpha(N-3)q}+1)
(e^{\theta_{ij}-\alpha(N-3)q}+1)+(e^{\alpha}-1)}\right\rangle_{\!\!\theta}}\\
& = {\left(\tfrac{1}{\sqrt{2\pi\sigma^2}}\right)}^3
			\iiint_{-\infty}^\infty
			\frac{e^{\alpha} \ e^{\frac{-(\theta_{ij}-\mu)^2}{2\sigma^2}} 
			\ e^{\frac{-(\theta_{jk}-\mu)^2}{2\sigma^2}} \ 
			e^{\frac{-(\theta_{ki}-\mu)^2}{2\sigma^2}}}{(e^{\theta_{jk}-\alpha(N-3)q}+1)
				(e^{\theta_{ki}-\alpha(N-3)q}+1)
				(e^{\theta_{ij}-\alpha(N-3)q}+1)+(e^{\alpha}-1)} 
				d\theta_{ij} \ d\theta_{jk} \ d\theta_{ki} .
\end{aligned}	
\end{equation}
Up to here, we presented an analytical solution for the Hamiltonian, 
Eq.~(\ref{eq2}), by the mean-field approach. We derived statistical 
quantities of the network that help us to study the behavior of the system 
on the average. In the following section, we simulate the system. The 
results show the consistency between analytical and simulation outcomes.
\end{widetext}

\section{\label{sec:res} Simulation}
The mean-field method that was studied in the previous section made us 
confirm our findings by simulation. For this purpose, we consider a 
network of size 
$N=50$ and investigate the behavior of the order parameter versus changing 
variance. In the current section, we explain details of simulation method 
based on the introduced Hamiltonian Eq.~\ref{eq2}. The network contains 
$N$ nodes and each link of the network can be either $0$ or $1$, so 
different configurations of links are 
possible and $2^{N}$ micro states are probable for the created network. 
Every micro state has specific energy depending on the number of 
links and triads of that configuration. This energy is equivalent to 
society stress in Heider's theory. One may consider a randomly chosen link 
and accept the flip of that link only when the next neighbor micro state 
has lower energy, to evolve the network through a path to lower stress. 
However, Antal \textit{et al.} have explained that sometimes the network 
gets trapped in local minima forever, and it can not find the global 
minimum anymore \cite{antal2005dynamics}. So, we let the network vary 
according to the Metropolis-Hastings algorithm to reach the global minimum 
of the energy landscape.
In the Metropolis algorithm, in each step, we choose a link randomly and 
calculate the total energy of the network before and after flipping the 
state 
of that link. We indicate the energy before and after changing as $E_1$ 
and $E_2$, so $\Delta E = E_2 - E_1$. If $\Delta E\le0$, the transition 
probability, $W_{1\rightarrow2}=1$ and this change will be accepted, 
otherwise ($\Delta E>0$) in the equilibrium state using the Boltzmann 
probability of being in the configuration $G$, $p(G)$, we can 
argue that
\begin{equation*}
	p(G_1)W_{1\rightarrow2} = p(G_2)W_{2\rightarrow1}.
\end{equation*}

Knowing that $p(G)=Z^{-1}e^{-H(G)}$ and $Z$ be the partition function that 
normalizes the probability distribution, and $W_{1\rightarrow2} = 
e^{-\Delta E}$ is 
transition probability of going from state $1$ to $2$. Actually in the 
Metropolis algorithm, the acceptance of transition even with an increase 
in the total energy is probable. We repeated Monte Carlo steps until the 
network reaches its equilibrium state. Then we calculated the values of 
statistical quantities $p = \langle 
 s_{ij}\rangle$, $q = \langle s_{ik} s_{kj} \rangle$, and
$r = \langle s_{ij} s_{jk}s_{ki} \rangle$ which are respectively average 
of link, two-star, and triad. Studying these quantities gives us 
remarkable information about the network structure. As the total energy of 
the network depends on the coefficient of triadic interactions, $\alpha$, 
and the fluctuations of the individual benefits $\theta_{ij}$, these two 
parameters indicate the evolution of the network.
We apply a range of the coefficient $\alpha$ and study the dynamics in the 
configurations. We change $\alpha$ from small values (\textit{i.e.}, weak 
interaction of 
triad) to large values and vice versa. Small values of $\alpha$ are proper 
for starting the simulation because the triadic interactions are weak 
against fluctuations and the system will easily reaches the equilibrium 
state and it won't be frozen in a local minimum. 
The important problem is how to set the $s_{ij}$ as initial conditions. 
Each $s_{ij}$ will take the value $1$ with the probability of $p_0$ and 
$0$ with the 
probability of $1-p_0$. The probability $p_0$ depends on the 
parameters $\bar{\theta}$ and $\sigma$. To find the suitable value of 
$p_0$ for a specific distribution of $\theta_{ij}$ and the small value of 
$\alpha$, we initially set the adjacency matrix with probability $0.5$, 
and we let the network reach the equilibrium state. Then we calculate 
the probability of the existence of the link in the equilibrium state 
which is the suitable value of $p_0$ for starting the simulation.
 
Then the simulation repeated for different values of $\bar{\theta}$, 
$\sigma$. We see that by increasing $\bar{\theta}$, the initial 
probability of existing links ($p_0$) will decrease and this is equivalent 
to sparsity in the network. In other words, considering bigger value for  
$\bar{\theta}$ corresponds to more terms of positive sign in the first 
summation of the Hamiltonian Eq.~(\ref{eq2}), following by forming more 
positive links and increment of energy which is not proper for the system, 
the system tends to omit those links and making the network more 
sparse. Finally, the quantities $p$, $q$ and $r$ are calculated for a 
range 
of $\alpha$. As Newman has shown, the behavior of these quantities with 
respect to the parameter $\alpha$ is similar \cite{Newman2005}; therefore, 
it suffices to present the behavior of the order parameter $q$ against 
$\alpha$ here.

\section{Heterogeneous Individual Benefit outcome}
\begin{figure}[ht] 
	\centering
	\subfigure[]{\label{theta1sigma0_001:Q}
		\hspace*{-1cm} 
		\includegraphics[height=60mm, trim={0 0cm 0 
			0cm},clip]{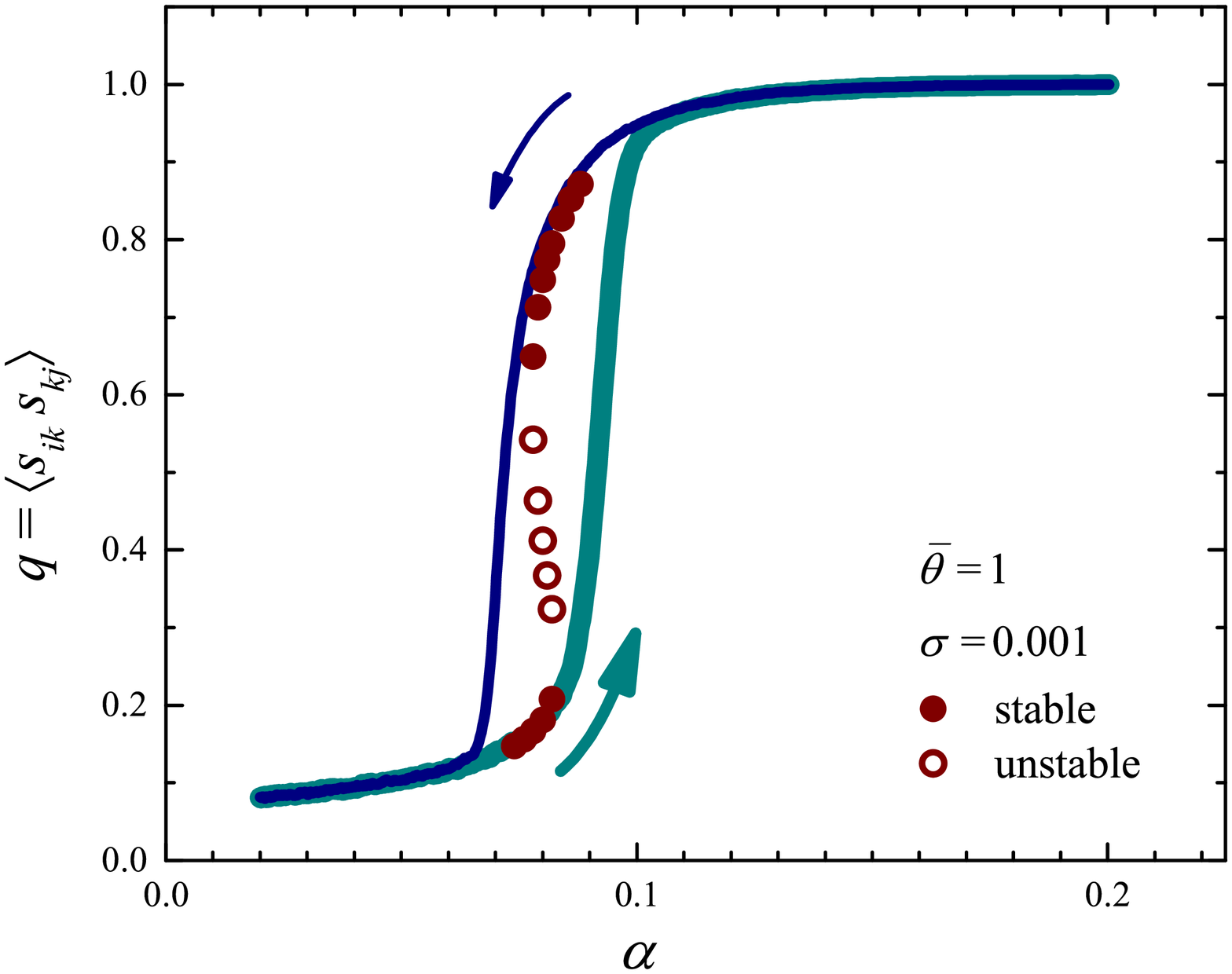}}	
	\subfigure[]{\label{theta0.54sig0.001:Q}\!\!\!\!\!\! 
		\includegraphics[height=30mm, trim={0 0cm 0 
			0cm},clip]{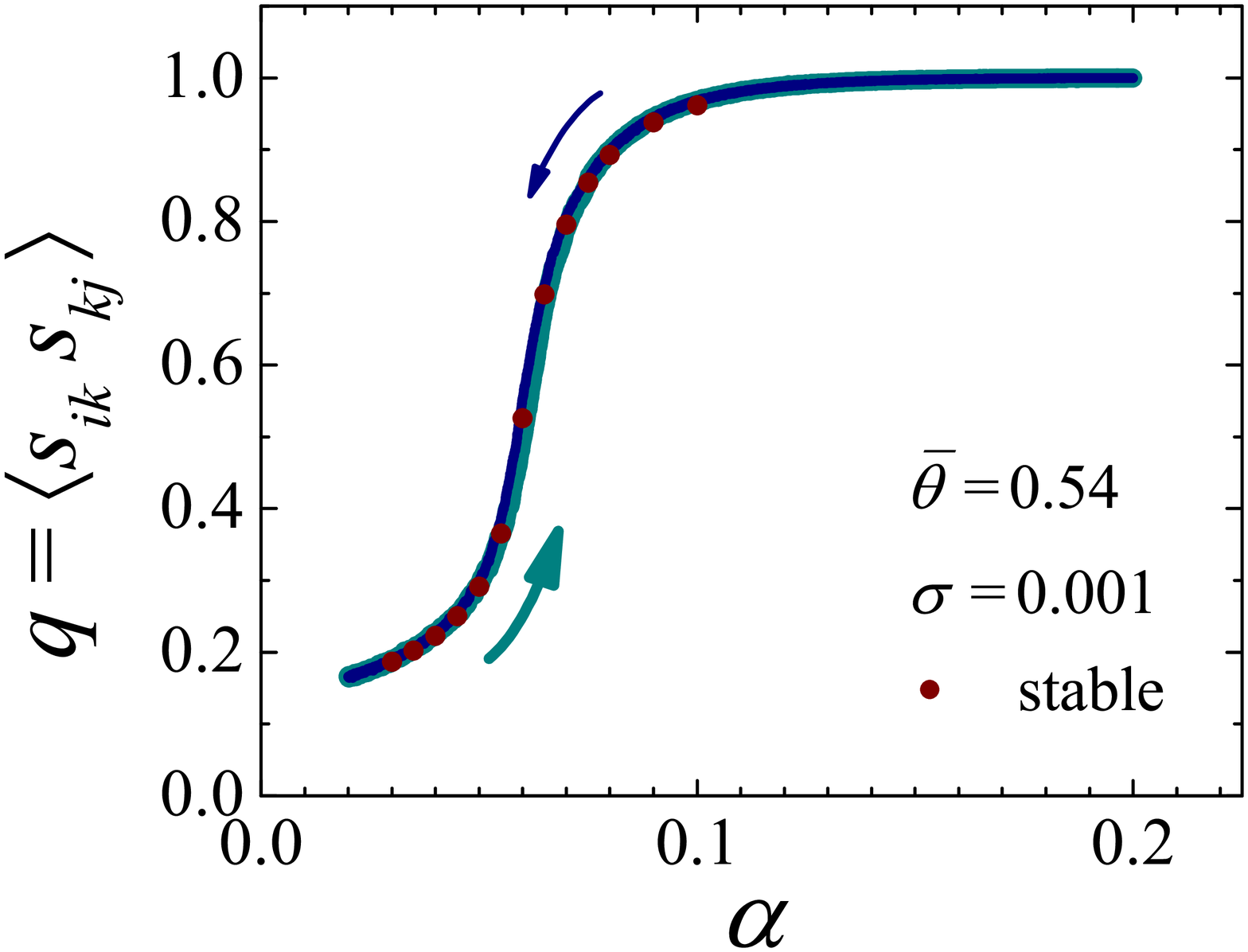}}
	\subfigure[]{\label{theta1sigma1:Q}
		\includegraphics[height=30mm, trim={0 0cm 0 
			0cm},clip]{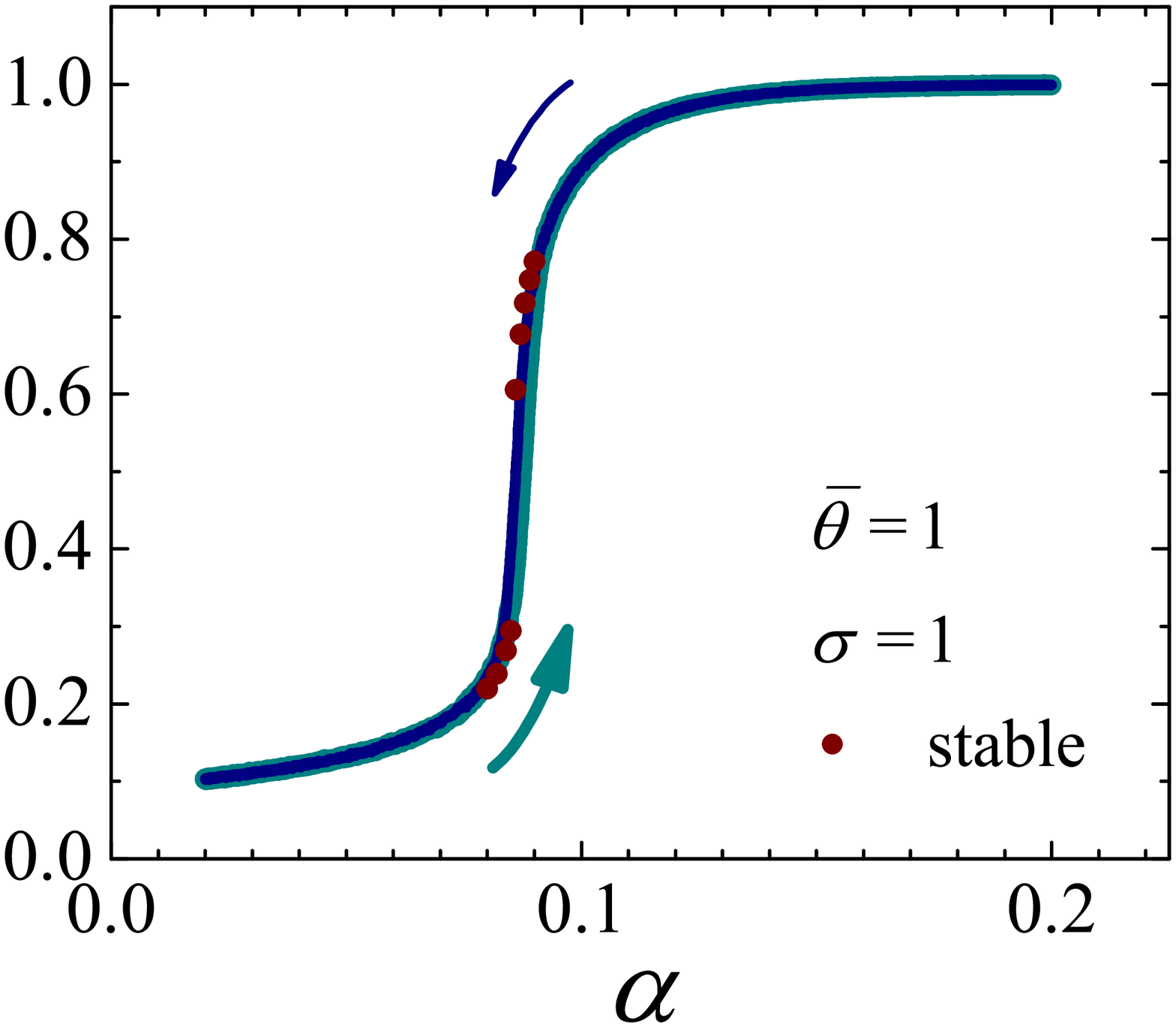}}
	\caption{\label{pqr10_001}
		Diagram of $q=\langle s_{ik}s_{kj} \rangle$ {\em vs.} $\alpha$ for 
		(a) $\bar\theta=1$ , (b) $\bar\theta=0.54$ and (c) $\bar\theta=1$. 
		In (a) and (b) graphs, $\sigma=0.001$, while in (c)
		$\sigma=1$ and 
		the network includes $50$ nodes. The solid 
		\textcolor{darkblue}{thin}/\textcolor{darkcyan}{\textbf{thick}} 
		curves show the results of the simulation while the solid/empty 
		circles represent the analytic results. In the simulation, to make 
		it easier to get to the ground state, the value of $\alpha$ was 
		first increased, then reduced from the final value to return to 
		its initial value (follow the arrows, $\arrows$). The solid 
		circles (\textcolor{darkred}{\tiny$\bullet$}) represent the stable 
		solution while the empty circles 
		(\textcolor{darkred}{\tiny$\boldsymbol{\circ}$}) denote the 
		unstable solution that the latter has not appeared in the 
		simulation results.}
\end{figure}

In this section, we compare the results of Monte Carlo simulation and 
analytical solution, and we see a good agreement between these two. The 
order parameter, $q$, was calculated in both 
approaches for different sets of parameters $\bar\theta$, $\sigma$. 
Obtained results are shown in 
Figs.~\ref{pqr10_001} for a range of $\alpha$. In the simulation, the 
order parameter $q$ was calculated in the equilibrium state.
In Figs.~\ref{pqr10_001}, the $q$ is increased as we increase the 
parameter $\alpha$. For small values of $\alpha$, the density of two-stars 
is low, and for large values of $\alpha$, it is close to $1$.
The existence of two regimes, the low and high density of two-stars, in 
our model is the consequence of competition between two parameters 
$\bar\theta$ and $\alpha$ in the Hamiltonian~(\ref{eq2}). For $\alpha \ll 
\bar\theta$, the winner of this competition is the first term, and it 
decreases the energy of the network by removing links. In this case, the 
mean value of links ($p=\langle s_{ij} \rangle$) will tend to $0$, and as 
a result, the mean value of two-stars will approximately go to zero. 
For $\alpha \gg \bar\theta$, the second term will be the winner, and as a 
result of creating triangles both the total energy minimizes and the 
average value of the two stars is raised to $1$.

In Fig.~\ref{pqr10_001}\subref{theta1sigma0_001:Q}, $q$ has rendered three 
answers for some value of $\alpha$, one of which is unstable as the two
others are stable. The appearance of the hysteresis 
loop means that the system stores its history and resists changes from the 
low to high density and vice versa. In this 
region, two stable solutions are probable but do not occur simultaneously 
and the system follows one of them. According to 
Fig.~\ref{pqr10_001}\subref{theta1sigma0_001:Q} starting from 
the small values of $\alpha$ and moving via lower 
branch, \textcolor{darkcyan}{\textbf{thick}} curve, it lasts for a while 
to make the transition. The same will happen by starting from big values 
of $\alpha$ and moving through upper branch, \textcolor{darkblue}{thin} 
curve. Therefore, the transition point in lower branch is different from 
that of the upper branch and a hysteresis loop appears. In order to 
illustrate how heterogeneity matters in the evolution of the network, we 
repeated analytical calculations and simulations for two cases of changing 
the mean value and the variance of Gaussian probability distribution of 
links, respectively in Fig.~\ref{pqr10_001}\subref{theta0.54sig0.001:Q}, 
and \subref{theta1sigma1:Q}.
Fig.~\ref{pqr10_001}\subref{theta0.54sig0.001:Q} 
displays that small quenched fluctuations in $\{\theta_{ij}\}$ confirms 
the results of Ref.~\cite{Newman2005}. The order parameter, $q$, has 
rendered only one answer for each particular value of $\alpha$ and 
there is no hysteresis loop. By increasing the variance 
of the Gaussian probability distribution of links which means inducing 
more 
heterogeneity in local benefits, the resistance of the system lessens 
hence, the two branches overlap 
Fig.~\ref{pqr10_001}\subref{theta1sigma1:Q}.
In the Monte Carlo 
simulations, as shown in the figures, only stable answers are obtained, 
while both the unstable and the stable answers are obtained by analytical 
solution. 

\begin{figure}[H]  
	\centering\includegraphics[width=0.9\columnwidth]{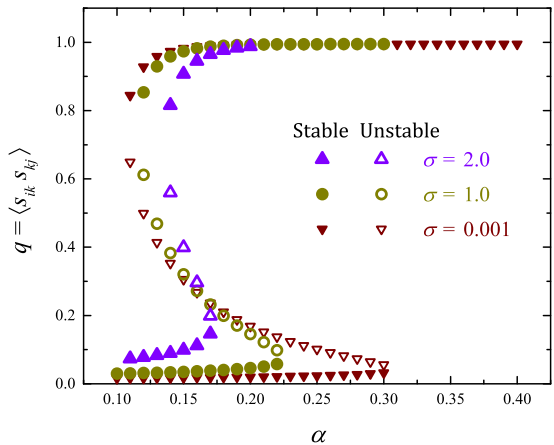}
	\caption{\label{theta2dffrntSigma} 
		Solid symbols show the stable analytical results for the average 
		of two-star $q$ {\em vs.} the coefficient $\alpha$ for the fixed 
		value of $\bar\theta=2$ and different values of $\sigma$, while 
		the empty 
		symbols indicate the unstable solutions.}
\end{figure}

Fig.~\ref{theta2dffrntSigma} is a confirmation for the noticeable role of 
the variance on the hysteresis loop. It depicts the order parameter, $q$, 
for 
$\bar\theta=2$ and different values of $\sigma$. It displays for a fixed 
value of the $\bar\theta$ by changing the variance the unstable 
answers deforms. For a range of $\alpha$, where the 
unstable solution occupies a wider area, reflects the resistance of the 
system against changes by remembering its past. It takes a while for the 
system to make the transition. By applying the larger $\sigma$, we induce 
more heterogeneity therefore the resistance of the system decreases and 
the unstable answers grab a narrower area hence the hysteresis loop gets 
smaller.  

\begin{figure*}[ht] 
	\centering 
	\includegraphics[width=2\columnwidth]{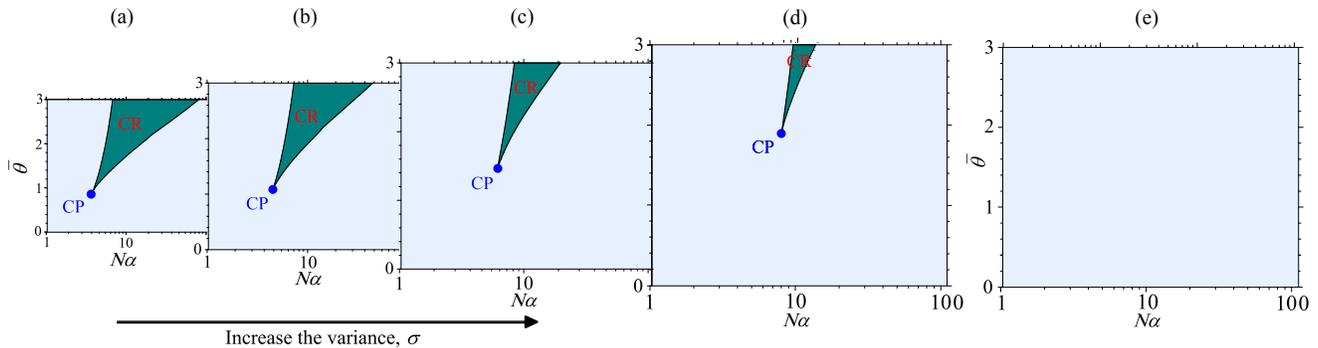}
	\caption{The phase diagram in (N$\alpha$, $\bar\theta$) space. For 
	figure a-e the variance of the Gaussian probability distribution of 
	individual benefit is as $0.001, 1, 2, 3, 10$ respectively.  
	The dark
		shaded area corresponds to the coexistence region $(CR)$ in which 
		the system can be in either of two stable states within a 
		hysteresis loop, one of high density (for large value of $\alpha$) 
		and one of low density of 
		triangles (for small value of $\alpha$). Increasing the variance 
		which is equal to inducing more 
		heterogeneity in the network result in shifting the critical point 
		$(CP)$ and the smaller coexistence region.}
	\label{diagramphase} 
\end{figure*}

Fig.~\ref{diagramphase}, indicates the phase diagram of the system. 
In the coexistence region, $CR$, (shaded area) the self-consistence 
equation 
(\ref{eq8}) has three solutions corresponds to a symmetric broken 
phase. In approval of what Newman stated in Ref \cite{Newman2005} there 
is a 
critical point, $CP$, separating the coexistence region from high 
symmetry and low symmetry region. Fig.~\ref{diagramphase} demonstrates the 
effect of the 
variance on the area of the shaded region and where its lowest point is 
the critical point. Notice that inducing more heterogeneity in the network 
makes the coexistence region narrower and change the position of the 
critical point below which the system makes a transition to second order 
phase transition phenomenologically \cite{Newman2005}.

\section{Conclusion}
Our relationships dictate that each person has an individual identity in 
addition to a social one and the individual benefit also plays a crucial 
role in the dynamics of social relationships. For this purpose complex 
networks represent explanatory models that illustrate us a general picture 
and lead us in direction of studying real social network. In many studies, 
the focus was more on global benefit, while in the Strauss model, the 
individual benefit was given importance along with global ones. Here we 
made a modification to the Strauss'model by taking into account the 
heterogeneity  of local benefits. Results show:
\begin{itemize}
	\item The system experiences a first-order phase transition and the 
	hysteresis loop is an indicator that shows the system remembers its 
	past.
	\item The existence of the hysteresis loop depends on unstable 
	solutions and increasing the variance of the Gaussian probability 
	distribution of links equals heterogeneity in local benefits result in 
	deformation of coexistence region and the width of the hysteresis loop 
	gets narrower.
	\item The phase diagram indicates that for a small value of $\alpha$ 
	system is in low density of triangles state and by increasing value of 
	$\alpha$ more triads forms and the system is in a condensed state of 
	triangles.
	\item The location of the critical point ($CP$) in the phase 
	diagram, depends on the mean and the
	variance of the Gaussian probability distribution of induced 
	heterogeneity and it is increasing by inducing more heterogeneity in 
	the network. 
\end{itemize}
\bibliography{MyReferences}
\end{document}